\documentclass[a4paper]{article}
\usepackage{geometry}

\usepackage{setspace}
\usepackage{float}
\usepackage{epstopdf}
\usepackage{standalone}
\usepackage{amsmath,amsfonts,amssymb}
\usepackage{color}

\usepackage{bm}
\usepackage{pdflscape}
\usepackage{authblk}
\usepackage{graphicx}
\usepackage{subcaption}
\usepackage[flushleft]{threeparttable}
\usepackage[comma,authoryear]{natbib}

\usepackage{mathtools}

\newcommand{\interior}[1]{%
 {\kern0pt#1}^{\mathrm{o}}%
}
\usepackage [english]{babel}
\usepackage [autostyle, english = american]{csquotes}
\MakeOuterQuote{"}

\usepackage{algorithm,algpseudocode}
\usepackage{caption}
\usepackage[makeroom]{cancel}
\newcommand{\mbR}{{\mathbb R}}

\makeatletter
\newcommand*\bigcdot{\mathpalette\bigcdot@{.5}}
\newcommand*\bigcdot@[2]{\mathbin{\vcenter{\hbox{\scalebox{#2}{$\m@th#1\bullet$}}}}}
\makeatother

\title{\textbf{An Interaction Neyman-Scott Point Process Model for Coronavirus Disease-19}}

\author[1,2]{Jaewoo Park}
\author[3]{Won Chang}
\author[4]{Boseung Choi}
\affil[1]{Department of Statistics and Data Science, Yonsei University}
\affil[2]{Department of Applied Statistics, Yonsei University}
\affil[3]{Division of Statistics and Data Science,  University of Cincinnati}
\affil[4]{Division of Big Data Science, Korea University}

\begin{document}

\maketitle

\begin{abstract}
With rapid transmission, the coronavirus disease 2019 (COVID-19) has led to over 2 million deaths worldwide, posing significant societal challenges. Understanding the spatial patterns of patient visits and detecting the local spreading events are crucial to controlling disease outbreaks. We analyze highly detailed COVID-19 contact tracing data collected from Seoul, which provides a unique opportunity to understand the mechanism of patient visit occurrence. Analyzing contact tracing data is challenging because patient visits show strong clustering patterns while clusters of events may have complex interaction behavior. To account for such behaviors, we develop a novel interaction Neyman-Scott process that regards the observed patient visit events as offsprings generated from a parent spreading event. Inference for such models is complicated since the likelihood involves intractable normalizing functions. To address this issue, we embed an auxiliary variable algorithm into our Markov chain Monte Carlo.  We fit our model to several simulated and real data examples under different outbreak scenarios and show that our method can describe spatial patterns of patient visits well. We also provide visualization tools that can inform public health interventions for infectious diseases such as social distancing.
\end{abstract}

\noindent%


\section{Introduction}

Caused by the transmission of severe acute respiratory syndrome coronavirus 2 (SARS-CoV-2), the coronavirus disease 2019 (COVID-19) was first reported in December 2019 in Wuhan, Hubei providence, China \citep{who2020coronavirus}. By February 2021, there have been 100 million confirmed cases of COVID-19, with more than 2 million deaths. The disease spreads more quickly than influenza, mainly through close contact with infected people \citep{cdc2020coronaspread}. Human-to-human transmission is most common for COVID-19, primarily via respiratory droplets or aerosols from an infected person. Thus contact tracing, which records the travel paths of confirmed patients in detail, is a highly effective disease control measure. Spatial point process models for contact tracing data can provide useful insights into the mechanism of patient visits. Here, we propose a new point process model for studying the probabilistic mechanism of patient visits. Our model can provide useful epidemiological information such as a warning system for local hotspots. 

Here we analyze the locations visited by confirmed patients in Seoul, South Korea. The data set contains full contact histories of confirmed patients, and is hence a unique source for point process modeling. Contact tracing data for most other countries contain only partial or imperfect tracing records. We regard each visit in the contact tracing data as an event in the point process model. Since such events are spatially clustered, a natural way to model this is the Neyman-Scott point process \citep{neyman1952theory}, which considers observed events as offsprings generated around unobserved parents. In the epidemiological context, we can regard a parent point as a cluster center of disease (i.e., spreading event), and offsprings are corresponding clusters (i.e., patient visits). 

Several computational methods have been developed for inference for point processes, including a minimum contrast method \citep{diggle2013statistical} and pseudolikelihood approximation \citep{guan2006composite, diggle2010partial}. However, such approaches are sensitive to the choice of tuning parameters and may not be accurate in the presence of strong spatial dependence among points. Through simulation studies, \cite{mrkvivcka2014two} reports that Bayesian estimation is the most precise for the Neyman-Scott point process. Furthermore, we can easily adjust priors based on epidemiological knowledge. Therefore, the Bayesian framework can be practical for fitting hierarchical point process models. 

A simple Neyman-Scott process has limited applicability here because it assumes an independent parent process.  Local spreading events (parents) may have complex interaction behavior, leading to an intractable likelihood function; this is common in many spatial point process models. Several extensions for Neyman-Scott processes have been developed. \cite{waagepetersen2007estimating, mrkvivcka2014two, mrkvivcka2017parameter} study inference for inhomogeneous Neyman-Scott point processes. \cite{yau2012generalization} considers a special case where the parent process exhibits repulsion. \cite{albert2019hierarchical} generalizes the Neyman–Scott process by allowing the parents to follow the log-Gaussian Cox process \citep{moller1998log}. However, these existing models may not be appropriate for our COVID-19 data because of the complex interactions of local spreading events. Unobserved spreading events attract each other at mid-range distances because other spreading events likely to appear in nearby communities. At the same time, spreading events repel each other at small distances to avoid merging. Otherwise, one spreading event would become offsprings to others. 

In this manuscript, we propose a novel parametric approach that exhibits interaction between unobserved spreading events. This new approach can infer model parameters that describe spatial patterns of spreading events. Our model allows us to detect unobserved spreading events, thereby providing a practical warning system for coronavirus hotspots. Recently, \cite{goldstein2014attraction} proposes an attraction-repulsion point process model by extending the Strauss process \citep{strauss1975model} that only allows repulsion behavior. \cite{russell2016dynamic} uses such a process to model different types of interactions among animals, including herding behavior and collision avoidance. In this manuscript, we develop an interaction Neyman-Scott point process to describe this behavior. We incorporate the attraction-repulsion \citep{goldstein2014attraction} process into the parent process. Our model includes an intractable normalizing function, posing computational and inferential challenges. To address this, we adopt a novel auxiliary variable Markov chain Monte Carlo (MCMC) \citep{moller2006efficient,liang2010double}, that can avoid the direct evaluation of intractable normalizing functions.

Note that susceptible-infected-recovered (SIR) models \citep{kermack1927contribution, dietz1967epidemics} are also popular for modeling infectious disease dynamics. There is a vast literature on modeling the COVID-19 data based on SIR type models \citep{choi2020estimating, fanelli2020analysis, anastassopoulou2020data, barlow2020accurate}. These models are typically based on daily counts of infection, death, and recovery. Rather than modeling confirmed cases, as in SIR models, we fit our point process model to locations visited by confirmed patients from our contact tracing data. Our approach can provide epidemiological insights for social distancing in daily life by using the full information of visiting history.

The remainder of this paper is as follows. In Section 2, we describe the COVID-19 contact tracing data in Seoul, South Korea. In Section 3, we introduce a new interaction Neyman-Scott process model and discuss its computational and inferential challenges. We describe our MCMC algorithm for this model and provide implementation details. In Section 4, we apply our methods to both simulated and real COVID-19 data sets. We show that our methods can detect the sources of spreading events and describe the spatial patterns of patient visits. Furthermore, we provide a disease risk map that can give important epidemiological interpretations. In Section 5, we conclude the paper with a discussion and summary.

\section{Data Description and Exploratory Analysis}

In this section, we provide the background of our contact tracing data and summarize the results from exploratory data analysis.

\subsection{COVID-19 Contact Tracing Data}

\label{sec:data_description}

We study the COVID-19 contacting tracing data in Seoul. During the early stages of the pandemic in South Korea, Daegu and the North Gyeongsang (NGO) province were the central areas of the disease outbreak. However, the regional government only provided data on daily new infections, deaths, and recoveries, rather than disclosing full contact tracing information. On the other hand, confirmed COVID-19 patients in Seoul were perfectly under control, with their full contact information recorded. When a patient is confirmed to be infected by a screening clinic, he or she is immediately quarantined. In addition, the local government tracks the patient's moving paths and posts the routes on local government websites. This information is legally disclosed for two weeks and then archived. This code follows the "Response Guidelines to Prevent the Spread of COVID-19 (local government)" \citep{guidelines2020}. The contact tracing data sets have been collected directly from these websites.

\begin{table}[tt]
\small
\centering
\begin{tabular}{ccccccc}
  \hline
  PATID & Date time & Address & Trans & Description & Latitude & Longitude \\
  \hline
XXX & 	0307 & 	Chunhodaero 145, Seoul & 	car & 	Screening Clinic & 	959371 & 	1952879 \\
XXX & 	0302 & 	Gyungheedaero 7-1,Seoul & 	other & Café &  	960469 & 	1954862 \\
XXX & 	0302 & 	HanChunro 58, Seoul & 	car &   	House & 	962047 & 	1956113 \\
XXX & 	0301 & 	Hoegiro 19, Seoul & 	other & 	Café & 	    960541 & 	1954855 \\
XXX & 	0301 & 	Hoegiro 18, Seoul & 	route & 	Restaurant & 960500 & 	1954704 \\
XXX & 	0301 & 	Imunro 37, Seoul & 	    route & 	Café & 	    960812 & 	1954692 \\
XXX & 	0301 & 	HanChunro 58, Seoul & 	other & 	House & 	962047 & 	1956113 \\
XXX & 	0229 & 	Shinimunro 40, Seoul & 	route & 	Hosipital & 961735 & 	1956027 \\
XXX & 	0229 & 	Shinimunro 24, Seoul & 	route & 	Pharmacy & 	961718 & 	1956025 \\
XXX & 	0229 & 	Hoegiro 21, Seoul & 	route & 	Restaurant & 960558 & 	1954887 \\
XXX & 	0229 & 	Imunro 37, Seoul & 	    car & 	    Café & 	    960812 & 	1954692 \\
XXX & 	0229 & 	Hwigyeongro 2, Seoul & 	route & 	Café & 	    961181 & 	1955099 \\
XXX & 	0229 & 	HanChunro 58, Seoul & 	other & 	House & 	962047 & 	1956113 \\
\hline
\end{tabular}
\caption{Movement path of a confirmed patient. For confidentiality, we do not report the patient's ID (PATID).}
\label{route} 
\end{table}

\begin{figure}[tt]
\begin{center}
\includegraphics[scale = 0.44]{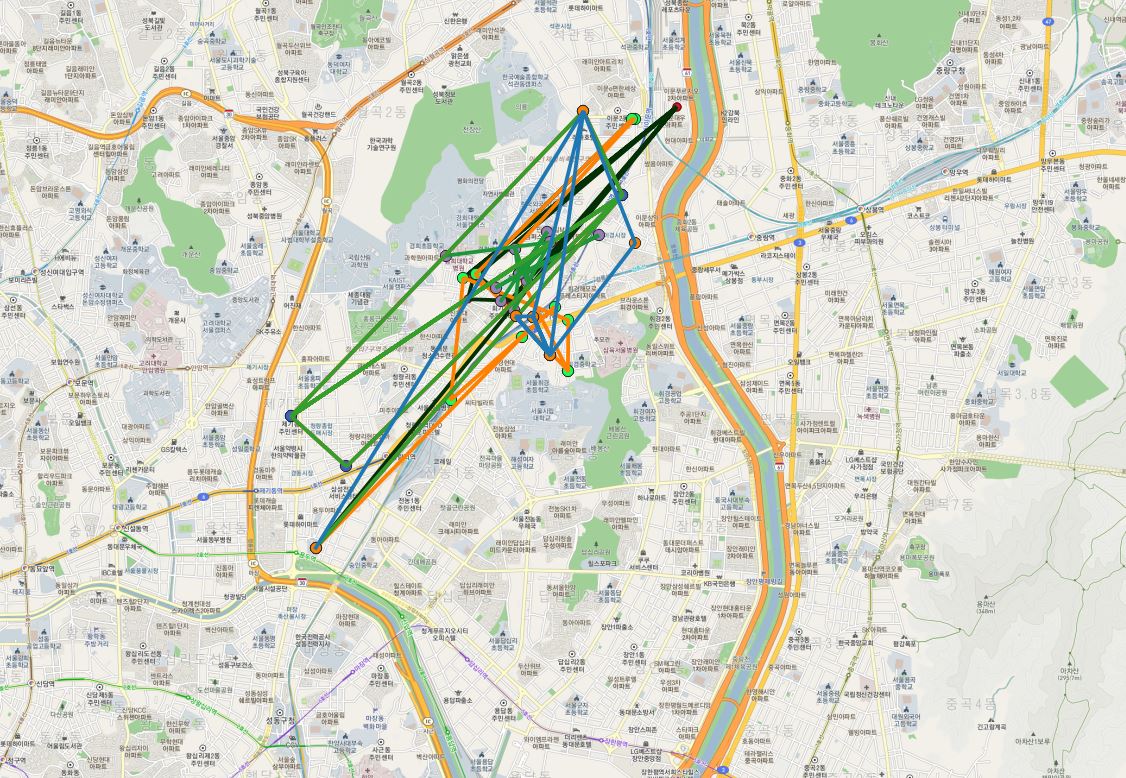}
\end{center}
\caption[]{Travel routes of five confirmed patients in the Dondaemoon-gu area. Each color represents a patient's path.}
\label{route2}
\end{figure}

Table~\ref{route} provides examples of contact tracing data for a patient. For each individual, the information for travel route (transportation, places visited, coordinates) has been recorded. For example, the patient in Table~\ref{route}, visited 13 places from February 29 to March 7, 2020. This patient was confirmed infected at a screening clinic in Seoul and immediately quarantined. Figure~\ref{route2} illustrates a graphical representation of the travel routes of five patients, including the patient in Table~\ref{route}. To study the spatial patterns of patient visits, we regard each visit (coordinate) as a realization of the point process.

In this paper, at any given time point, we fit a point process model based on contact tracing data accumulated over the last two weeks. According to the basic data analysis for the early stages of the COVID-19 spread in South Korea and China, two weeks is the mean period of recovery from infection \citep{ki2020epidemiologic,choi2020estimating}. Two weeks is also the period for epidemic investigation conducted by the local health authority for all confirmed cases. Therefore, when we construct a warning system for COVID-19 at a certain time, modeling patient visits in the last two weeks would be most useful. As illustration examples, we fit our model for non-overlapping 8 different time periods from February 20th to June 11th in 2020. Our goal here is to examine how our model works and provide important epidemiological insights for these different time periods with various disease spreading patterns. In particular, in the main document we focus on three consecutive time periods that end at March 19th, April 2nd, and April 15th which can be considered as severe, moderate, and mild outbreak cases, respectively in terms of overall visit numbers. The results for other time periods are provided in the supplementary material.

\subsection{Exploratory Data Analysis}

Spatial point processes provide a natural solution to model spatial patterns for locations visited by confirmed patients. Here, we provide the motivation for a new interaction point process model with some exploratory data analysis. Let $\mathbf{X}=\lbrace x_1,\cdots,x_n \rbrace$ be a realization of point process over the bounded spatial domain $\mathcal{S}\in \mbR^2$. The pair correlation function (PCF) is useful for exploring point process observations \citep{stoyan1994fractals}. The PCF is defined as
\[
J(r)=\frac{1}{2\pi r}K'(r),~~~K(r)=\frac{|\mathcal{S}|}{n^2}E\Big[\sum_{i \neq j}1_{\|x_i-x_j\|\leq r} \Big], 
\]
where $|\mathcal{S}|$ is the area of the spatial domain. Here, Ripley's $K(r)$ is the expected density of points within distance $r$. Under complete spatial randomness, $K(r)=\pi r^2$, which results in $J(r)=1$. $J(r)>1$ indicates that points have a tendency to cluster at distance $r$ (attraction) while $J(r)<1$ indicates that points tend to remain apart at distance $r$ (repulsion). 

\begin{figure}[tt]
\begin{center}
\includegraphics[scale = 0.52]{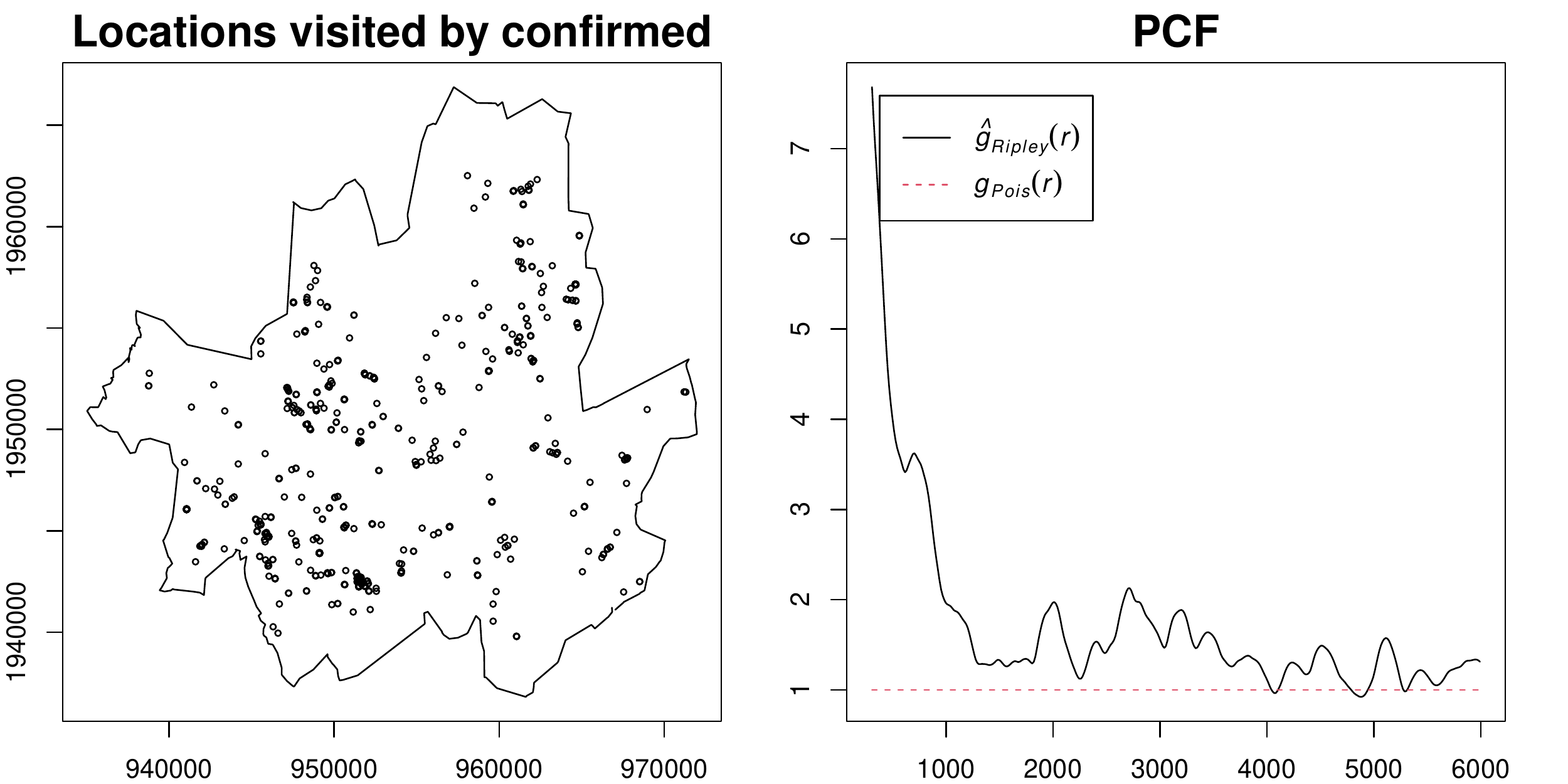}
\end{center}
\caption[]{The left panel shows the locations visited by patients (March 6 - March 19, 2020). The right panel shows the estimated PCF from the observation. The red line indicates theoretical PCF under complete spatial randomness.}
\label{edapoint}
\end{figure}

Figure~\ref{edapoint} illustrates an example of COVID-19 data and their PCF. We observe that patient visits are spatially clustered ($J(r)>1$). Therefore, a point process model for COVID-19 should capture such behavior. The Neyman-Scott process \citep{neyman1952theory} is widely used to study spatially aggregated point patterns. Furthermore, the Neyman-Scott process can detect cluster centers, which can be regarded as spreading events in our application; identifying the spreading center of the disease is crucial in our problem. Consider the Neyman-Scott process $\mathbf{X}=\cup_{c \in \mathbf{C}}\mathbf{X}_{c}$, where $\mathbf{X}_{c}$ is the offspring (clusters) and $\mathbf{C}$ is the parent (cluster centers). Given parent process $\mathbf{C}$, offspring $\mathbf{X}_{c},c \in \mathbf{C}$ follows an independent Poisson process with intensity $\alpha k(u-c,\omega)$. Here, $\omega$ controls the spread of offsprings around their parent, and $\alpha$ determines the expected number of offsprings per each cluster.  With the  Gaussian kernel $k(u-c,\omega)$ with mean $c$ and variance $\omega^2$, $\mathbf{X}$ is called the Thomas process \citep{thomas1949generalization}. The basic Neyman-Scott process models the unobserved parent process $\mathbf{C}$ as a simple Poisson process. 

The Neyman-Scott process is appropriate for modeling COVID-19 data because the locations visited by confirmed patients, $\mathbf{X}_c$, are clustered around the unobserved spreading event $\mathbf{C}$. From this, we can detect the local spreading events (cluster center of patient visits) of COVID-19. However, local spreading events $\mathbf{C}$ may have complex interactions. At mid-range distances $r$, spreading events tend to clump together because other spreading events likely to exist in the nearby region. At small $r$, spreading events tend to remain apart; otherwise, they would become offsprings of other spreading events, and hence two clusters can be `merged.' The basic Neyman-Scott process cannot describe such behavior because the basic model considers an independent Poisson process for $\mathbf{C}$. In the following section, we add another layer of complexity to the basic Neyman-Scott process. This new model can provide epidemiological interpretation for understanding spreading events for COVID-19.

\section{An Interaction Neyman-Scott Point Process Model}

In this section, we propose a new Neyman-Scott process by incorporating interaction behavior into $\mathbf{C}$. This new parametric approach can detect unobserved spreading events of COVID-19 and describe their patterns. Based on this, we can provide a warning system for higher risk regions. 

\subsection{Model}

Consider the realization of Neyman-Scott point process $\mathbf{X}=\lbrace x_1,\cdots,x_n\rbrace$ in the bounded plane $\mathcal{S} \subset \mbR^2$ (Seoul domain). This indicates the observed locations visited by the confirmed patients. Let $\mathbf{C}=\lbrace c_1,\cdots,c_m\rbrace$ be their unobserved parent process. In our context, each $c_i$ is the location of a spreading event in a local community. Note that Seoul has a two-level hierarchical administrative structure, Gu-Dong: Gu consists of multiple Dongs and each Dong has average size of $1.427$ km$^2$. We focus on modeling spreading events with a size of Dong because this coincides with most citizens' life radius. From this, we can provide a warning system for distancing in daily life. Spreading events tend to remain apart (repulsion) at small distances to avoid merging; otherwise, they would become offsprings to each other. At mid-range distances, they tend to clump together (attraction), because other spreading events likely to appear in nearby areas. Spreading events become independent at sufficiently large distances. To describe such patterns, we model the parent in the Neyman-Scott process as an attraction-repulsion interaction point process \citep{goldstein2014attraction}. The locations of unobserved spreading events (parent process) is modeled as 
\begin{equation}
f(\mathbf{C}|\kappa,\theta_{1},\theta_{2}) = \frac{\kappa^{m}\prod_{i=1}^{m}  \exp\left\lbrace \min\left(\sum_{i\neq j}\log{(\phi(D_{i,j}))},2\right)  \right\rbrace }{Z(\kappa,\theta_{1},\theta_{2})},
\label{Clik}
\end{equation}
where the interaction function is 
\begin{equation}
\phi(D) = \begin{cases}
      \theta_{1}-\left(\frac{\sqrt{\theta_{1}}}{\theta_{2}}(D-\theta_{2}) \right)^{2}  & 0< D \leq D_{1} \\
      1+\frac{1}{(0.5(D-D_{2}))^{2}} & D > D_{1}
\end{cases}
\label{interaction}
\end{equation}
Here, $D_1, D_2$ are set to make the interaction function $\phi(D)$ continuously differentiable \citep{goldstein2014attraction}. The interaction function $\phi(D)$ is determined by the distance between two parent points $c_i$, $c_j$. There are three parameters $\lbrace \kappa, \theta_1, \theta_2 \rbrace$ in this model: $\kappa$ controls the overall intensity of the parent process and $\lbrace \theta_1, \theta_2\rbrace$ control the shape of the interaction function $\phi(D)$. $\theta_1$ gives the peak value of $\phi$, whereas $\theta_2$ gives the location of the peak value. When the distance between spreading events $D_{ij}$ becomes too small, $\phi$ is less than 1, which means spreading events show repulsion behavior. On the other hand, as $D_{ij}$ becomes larger, $\phi$ is increased, which means spreading events show attraction behavior at mid-range distance. Such attractions become weaker as the distance between the spreading events becomes larger. From this, \eqref{Clik} can describe the attraction-repulsion behavior of local spreading events. 

Given the local spreading events, the locations visited by the confirmed patients can be modeled as 
\begin{equation}
f(\mathbf{X}|\mathbf{C},\alpha,\omega) = \exp\Big(|\mathcal{S}|-\int_\mathcal{S}\sum_{i=1}^{m} \alpha k(u-c_i,\omega)du \Big)\prod_{j=1}^{n}\sum_{i=1}^{m} \alpha k(x_j-c_i,\omega).
\label{Xlik}
\end{equation}
We use a symmetric Gaussian kernel with a center at each spreading event $c_i$ and variance $\omega^2$. This results in a higher intensity of patient visits around the spreading events. In our context, $\alpha$ controls the expected number of confirmed patients for each spreading event, and $\omega$ controls the levels of spreading events activity.

\subsection{Computational and Inferential Challenges}

The proposed model in the previous section has a hierarchical structure. Unobserved local spreading events (parent process) follow the spatial interaction process defined in \eqref{Clik} and \eqref{interaction}. Observed patient visits (offsprings) are distributed around the unobserved parents with Gaussian kernels. The Bayesian framework is useful for such hierarchical models in that it can easily quantify the model parameters' uncertainty using MCMC. With priors $p(\bm{\Theta})$, the joint posterior distribution is 
\begin{equation}
 \pi(\bm{\Theta}, \mathbf{C}|\mathbf{X}) \propto 
f(\mathbf{X}|\mathbf{C},\alpha,\omega)f(\mathbf{C}|\kappa,\theta_{1},\theta_{2})p(\bm{\Theta}),~~~\bm{\Theta} = \lbrace \alpha, \omega, \kappa, \theta_{1}, \theta_{2} \rbrace.
\label{fullmodel}
\end{equation}
Although \eqref{fullmodel} is a natural way to construct a joint posterior distribution, there are computational and inferential challenges to fit such a model due to the intractable normalizing functions. The model for spreading events in \eqref{Clik} can be written as 
\begin{equation}
f(\mathbf{C}|\kappa,\theta_{1},\theta_{2}) = \frac{h(\mathbf{C}|\kappa,\theta_{1},\theta_{2})}{Z(\kappa,\theta_{1},\theta_{2})} = \frac{\kappa^{m}\prod_{i=1}^{m}  \exp\left\lbrace \min\left(\sum_{i\neq j}\log{(\phi(D_{i,j}))},2\right)  \right\rbrace }{Z(\kappa,\theta_{1},\theta_{2})}.
\label{Clik2}
\end{equation}
Here, the calculation of normalizing function $Z(\kappa,\theta_{1},\theta_{2})$ requires integration over the spatial domain $\mathcal{S}$, which is infeasible. Inference for such models is challenging because an intractable $Z(\kappa,\theta_{1},\theta_{2})$ is a function of model parameters. The resulting posterior \eqref{fullmodel} is referred to as a doubly-intractable distribution \citep{murray2006}.

Several computational methods have been proposed to address this issue. By assuming conditional independence of points, \cite{besag1974spatial} proposed the pseudolikelihood that does not have intractable normalizing functions. Due to its convenience, pseudolikelihood approximations are often used in many point process applications (e.g., \citealp{diggle2010partial,tamayo2014modelling}). However, it is well known that such approximations are unreliable when there is strong dependence among points \citep{mrkvivcka2014two}. \cite{geyer1992constrained} propose theoretically justified Monte Carlo maximum likelihood methods, which maximize a Monte Carlo approximation to the likelihood. However, this has limited applicability because the algorithm requires gradients for $h(\mathbf{C}|\kappa,\theta_1,\theta_2)$, which is not available for interaction point process models such as the one defined above. Bayesian approaches can be an alternative for such cases. Several Bayesian methods have been proposed for doubly-intractable distribution (see \cite{park2018bayesian} for a comprehensive review). Among current approaches, \cite{park2018bayesian} reports that double Metropolis-Hastings (DMH) is the most practical method for point process models. DMH can avoid the calculation of $Z(\kappa,\theta_1,\theta_2)$ and alleviate memory issues, which can be serious for adaptive algorithms \citep{atchade2008bayesian, liang2015adaptive}. In the following section, we formulate DMH that can carry out Bayesian inference for an interaction Neyman-Scott point process model.

\subsection{Markov Chain Monte Carlo}

Here, we describe an MCMC algorithm for the attraction-repulsion Neyman-Scott point process model. Consider the model parameters $\bm{\Theta}^{(t)} = \lbrace \alpha^{(t)}, \omega^{(t)}, \kappa^{(t)}, \theta_{1}^{(t)}, \theta_{2}^{(t)} \rbrace$ and latent parent process $\mathbf{C}^{(t)}$ in \eqref{fullmodel} at the $t$-th iteration. We update the parameters successively. Offspring parameters $\alpha^{(t+1)}, \omega^{(t+1)}$ can be updated from 
\[
\alpha^{(t+1)}, \omega^{(t+1)} \sim f(\mathbf{X}|\mathbf{C}^{(t)},\alpha^{(t)},\omega^{(t)})p(\alpha^{(t)},\omega^{(t)})
\]
with a joint prior density $p(\alpha^{(t)},\omega^{(t)})$. We can update parent parameters from 
\[
\kappa^{(t+1)},\theta_{1}^{(t+1)},\theta_{2}^{(t+1)} \sim f(\mathbf{C}^{(t)}|\kappa^{(t)},\theta_{1}^{(t)},\theta_{2}^{(t)})p(\kappa^{(t)},\theta_{1}^{(t)},\theta_{2}^{(t)})
\]
with a joint prior density $p(\kappa^{(t)},\theta_{1}^{(t)},\theta_{2}^{(t)})$ (see Section \ref{prior} for how we choose the priors in our problem). Compared to updating offspring parameters, updating parent parameters is challenging due to the intractable normalizing function in \eqref{Clik2}. Intractable $Z(\kappa,\theta_1,\theta_2)$ leads to the intractable acceptance probability in MCMC as
\begin{equation}
\alpha = \min\left\lbrace \frac{  \frac{  h(\mathbf{C}^{(t)}|\kappa',\theta^{'}_{1},\theta^{'}_{2})  }{ Z(\kappa^{'}_1,\theta^{'}_1,\theta^{'}_2)}  p(\kappa^{'},\theta^{'}_{1},\theta^{'}_{2})q(\kappa',\theta^{'}_{1},\theta^{'}_{2}|\kappa^{(t)},\theta^{(t)}_{1},\theta^{(t)}_{2})     }{  \frac{h(\mathbf{C}^{(t)}|\kappa^{(t)},\theta^{(t)}_{1},\theta^{(t)}_{2})}{Z(\kappa^{(t)}_1,\theta^{(t)}_1,\theta^{(t)}_2)}p(\kappa^{(t)},\theta^{(t)}_{1},\theta^{(t)}_{2})q(\kappa^{(t)},\theta^{(t)}_{1},\theta^{(t)}_{2}|\kappa',\theta^{'}_{1},\theta^{'}_{2})      }, 1 \right\rbrace
\label{MCMCalpha}
\end{equation}
for the proposed parameters $\kappa',\theta_{1}^{'}$, and $\theta_{2}^{'}$. To avoid direct evaluation of the intractable normalizing functions in $\alpha$, we incorporate double Metropolis-Hastings (DMH) \citep{liang2010double} into our MCMC algorithm. Instead of evaluating the intractable normalizing functions in the $\alpha$, DMH sampler generates an auxiliary variable $\mathbf{A}$ from \eqref{Clik} via a birth-death MCMC \citep{moller2003statistical}. We provide details of the birth-death MCMC in the supplementary material. Note that the auxiliary variable $\mathbf{A}$ is from the same probability model as $\mathbf{C}$ and can be regarded as synthetic point process data.
The acceptance probability of DMH is now given as 
\begin{equation}
\alpha = \min\left\lbrace \frac{  h(\mathbf{C}^{(t)}|\kappa',\theta^{'}_{1},\theta^{'}_{2})h(\mathbf{A}|\kappa^{(t)},\theta^{(t)}_{1},\theta^{(t)}_{2})p(\kappa^{'},\theta^{'}_{1},\theta^{'}_{2})q(\kappa',\theta^{'}_{1},\theta^{'}_{2}|\kappa^{(t)},\theta^{(t)}_{1},\theta^{(t)}_{2})     }{  h(\mathbf{C}^{(t)}|\kappa^{(t)},\theta^{(t)}_{1},\theta^{(t)}_{2})h(\mathbf{A}|\kappa',\theta^{'}_{1},\theta^{'}_{2})p(\kappa^{(t)},\theta^{(t)}_{1},\theta^{(t)}_{2})q(\kappa^{(t)},\theta^{(t)}_{1},\theta^{(t)}_{2}|\kappa',\theta^{'}_{1},\theta^{'}_{2})      }, 1 \right\rbrace.
\label{DMHalpha}
\end{equation}
\noindent This does not include intractable terms because the introduction of $\mathbf{A}$ modifies the original densities in \eqref{MCMCalpha} by multiplying $h(\mathbf{A}|\kappa^{(t)},\theta^{(t)}_{1},\theta^{(t)}_{2})/Z(\kappa^{(t)},\theta^{(t)}_{1},\theta^{(t)}_{2})$ into the numerator and $h(\mathbf{A}|\kappa',\theta^{'}_{1},\theta^{'}_{2})/Z(\kappa',\theta^{'}_{1},\theta^{'}_{2})$ into the denominator. If the simulated auxiliary variable $\mathbf{A}$ resembles the parent process $\mathbf{C}$, the proposed parameters $\kappa',\theta_{1}^{'},\theta_{2}^{'}$ are likely to be accepted. 

Finally, we obtain $\mathbf{C}^{(t+1)}$ from 
\[
\mathbf{C}^{(t+1)} \sim f(\mathbf{X}|\mathbf{C}^{(t)},\alpha^{(t+1)},\omega^{(t+1)})f(\mathbf{C}^{(t)}|\kappa^{(t+1)},\theta_{1}^{(t+1)},\theta_{2}^{(t+1)})
\]
using a birth-death MCMC. Note that stationary distribution of the birth-death MCMC algorithm for updating $\mathbf{C}^{(t+1)}$ is different from that of generating the auxiliary variable $\mathbf{A}$ in DMH. The stationary distribution for generating the auxiliary variable follows \eqref{Clik} (see the supplementary material for details). Our MCMC algorithm is summarized in Algorithm~\ref{DMHalg}.

\begin{algorithm}[htbp]
\caption{MCMC for an attraction-repulsion Neyman-Scott process}\label{DMHalg}
\begin{algorithmic}
\normalsize
\State \textbf{Given $\Theta^{(t)}=\lbrace \alpha^{(t)},\omega^{(t)},\kappa^{(t)},\theta_{1}^{(t)},\theta_{2}^{(t)} \rbrace$ and $\mathbf{C}^{(t)}$ at $t$-th iteration. } \\ 

\State \textbf{Part 1: Update offspring parameters: $\alpha^{(t)}, \omega^{(t)}$. }\\

\State Propose $\alpha', \omega' \sim q(\cdot|\alpha^{(t)},\omega^{(t)})$ and accept it with probability\\
$$\alpha=\min\left\lbrace \frac{   f(\mathbf{X}|\mathbf{C}^{(t)},\alpha',\omega')p(\alpha',\omega')q(\alpha^{(t)},\omega^{(t)}|\alpha',\omega')   }{ f(\mathbf{X}|\mathbf{C}^{(t)},\alpha^{(t)},\omega^{(t)})p(\alpha^{(t)},\omega^{(t)})q(\alpha',\omega'|\alpha^{(t)},\omega^{(t)}) }, 1 \right\rbrace$$ else reject (set $\alpha^{(t+1)}=\alpha^{(t)}$, $\omega^{(t+1)}=\omega^{(t)}$ ).\\

\State \textbf{Part 2: Update parent parameters: $\kappa^{(t)}, \theta_{1}^{(t)}, \theta_{2}^{(t)}$. }\\

\State Propose $\kappa',\theta^{'}_{1},\theta^{'}_{2} \sim q(\cdot|\kappa^{(t)},\theta^{(t)}_{1},\theta^{(t)}_{2})$.\\
 
\State Generate an auxiliary variable $\mathbf{A}$ from the probability model using birth-death MCMC: \\
$\mathbf{A} \sim f(\mathbf{C}^{(t)}|\kappa',\theta^{'}_{1},\theta^{'}_{2})$.\\  

\State Accept it with probability
$$\alpha = \min\left\lbrace \frac{  h(\mathbf{C}^{(t)}|\kappa',\theta^{'}_{1},\theta^{'}_{2})h(\mathbf{A}|\kappa^{(t)},\theta^{(t)}_{1},\theta^{(t)}_{2})p(\kappa',\theta^{'}_{1},\theta^{'}_{2})q(\kappa',\theta^{'}_{1},\theta^{'}_{2}|\kappa^{(t)},\theta^{(t)}_{1},\theta^{(t)}_{2})     }{  h(\mathbf{C}^{(t)}|\kappa^{(t)},\theta^{(t)}_{1},\theta^{(t)}_{2})h(\mathbf{A}|\kappa',\theta^{'}_{1},\theta^{'}_{2})p(\kappa^{(t)},\theta^{(t)}_{1},\theta^{(t)}_{2})q(\kappa^{(t)},\theta^{(t)}_{1},\theta^{(t)}_{2}|\kappa',\theta^{'}_{1},\theta^{'}_{2})      }, 1 \right\rbrace$$ else reject (set $\kappa^{(t+1)}=\kappa^{(t)}, \theta^{(t+1)}_{1}=\theta^{(t)}_{1},\theta^{(t+1)}_{2}=\theta^{(t)}_{2}$).\\

\State \textbf{Part 3: Update unobserved parent process:  $\mathbf{C}^{(t)}$. }\\

\State $\mathbf{C}^{(t+1)} \sim f(\mathbf{X}|\mathbf{C}^{(t)},\alpha^{(t+1)},\omega^{(t+1)})f(\mathbf{C}^{(t)}| \kappa^{(t+1)},\theta^{(t+1)}_{1},\theta^{(t+1)}_{2})$ using birth-death MCMC

\end{algorithmic}
\end{algorithm}

\section{Application}

In this section, we apply our approach to simulated and real data examples. The computer code for this is implemented in {\tt R} and {\tt C++} using the {\tt Rcpp} and {\tt RcppArmadillo} packages \citep{eddelbuettel2011rcpp}.

\subsection{Prior Specification} \label{prior}

To complete the posterior specification we now explicitly define the prior distributions. In Neymann-Scott process, there is a strong dependence between $\alpha$ and $\omega$. The same observed pattern can be regarded as points generated from a small number of parents, with each of them has a large number of offsprings (large $\alpha$, large $\omega$), or they may come from a large number of parents, with each of them has a small number of offsprings (small $\alpha$, small $\omega$). To avoid such identifiability issues, we need to set the upper and lower bounds for them using uniform priors \citep{moller2007modern,kopecky2016bayesian}. 

As pointed out in Section 3.1, we focus on modeling spreading events in Dong scale. To this end we set lower and upper bound for $\omega$ based on the average size of a Dong. This encourages that patient visits from different Dongs come from different spreading events. Therefore, we set the range for $\omega$ to $[\sqrt{|\mathcal{S}|}/70, \sqrt{|\mathcal{S}|}/25]$. We assume that each spreading event (parent) generates at least 3 and at most 30 patient visits (offsprings) in a Dong on average. This choice of priors can avoid identifiability issues in the Neyman-Scott point process while maintaining interpretability and modeling flexibility. 

The prior range for $\kappa$ needs to be determined in consideration of the overall intensity and the offspring intensity $\alpha$. The overall intensity is around $6\times10^{-7}$, computed as dividing the average number of patient visits within two weeks by the area of Seoul $|\mathcal{S}|$. The average number of patient visits is computed over the 8 time periods. Since $\kappa$ controls for parent process intensity and $\alpha$ controls for average number of offsprings per each parent, $\kappa \times \alpha$ should be around the overall intensity $6\times10^{-7}$. By specifying the prior of $\kappa$ to be uniform with range $[1\times 10^{-10}, 1\times 10^{-6}]$, $\kappa \times \alpha$ belongs to the range $[3\times 10^{-10}, 3\times 10^{-5}]$ so that the resulting range can include the overall intensity $6\times10^{-7}$.

Similar to $\omega$, we set a uniform prior for $\theta_2$ whose upper and lower bounds are proportional to the average size of a Dong. We use a uniform prior with range $[\sqrt{|\mathcal{S}|}/70, \sqrt{|\mathcal{S}|}/25]$ so that  $\theta_2$ gives the location of the peak value within a Dong. To allow for attraction, we set the prior  for $\theta_1$ to be uniform with range $[1,3]$ as in  \cite{goldstein2014attraction}.

\subsection{Simulated Data}
We now conduct a simulation study to validate our method. We first simulate the parent process $\mathbf{C}$ from \eqref{Clik} with true parent parameters $\kappa,\theta_1$ and $\theta_2$. We then simulate offsprings $\mathbf{X}$ from \eqref{Xlik} centered around each parent point with true offspring parameters $\alpha$ and $\omega$. We consider three different scenarios here: In Scenario 1, we emulate COVID-19 data with a severe outbreak. This has numerous local spreading events ($m=129$) and each of them also generates a large number of offsprings. This results in $n=798$ patient visits in total. In Scenario 2, we consider a moderate outbreak. This has fewer spreading events ($m=93$) and each generates a moderate number of offsprings; the simulated patient visits consisted of $n=446$ points. Finally, in Scenario 3, we consider a mild outbreak in which we have the smallest number of spreading events ($m=56$) and patient visits ($n=239)$.

\begin{table}[tt]
\centering
\begin{tabular}{cccccc}
  \hline
  & $\alpha$ & $\omega$ & $\kappa \times 10^7$ & $\theta_{1}$ & $\theta_{2}$ \\
  \hline
Scenario 1  \\
Truth & 6 & 360 & 1.2 & 1.5 & 600\\
Estimates & 5.70 & 370.34 & 1.13 & 1.85 & 626.81  \\
95\%HPD & (4.79, 6.62) & (356.27, 390.39) & (0.39, 2.05) & (1.25, 2.67) & (500.02, 798.47)\\
\hline
Scenario 2  \\
Truth & 5 & 400 & 1.0 & 1.5 & 650\\
Estimates & 5.58 & 400.02 & 1.18 & 1.40 & 684.59 \\
95\%HPD & (4.72, 6.52) & (373.65, 427.80) & (0.60, 1.79) & (1.00, 1.93) & (500.02, 937.61) \\
\hline
Scenario 3  \\
Truth & 4 & 440 & 0.5 & 1.5 & 700\\
Estimates & 4.12 & 423.73 & 0.75 & 1.64 & 716.71\\
95\%HPD & (3.20, 5.02) & (387.68, 461.51) & (0.34, 1.22) & (1.03, 2.36) & (500.03, 941.41)\\
\hline
\end{tabular}
\caption{Inference results for simulated data sets. 100,000 MCMC samples are generated, discarding the first 50,000 as burn-in. Posterior mean estimates are reported for each parameter. The highest posterior density (HPD) is calculated using the {\tt coda} package.}
\label{simulest} 
\end{table}

\begin{figure}[hh]
\begin{center}
\includegraphics[scale = 0.6]{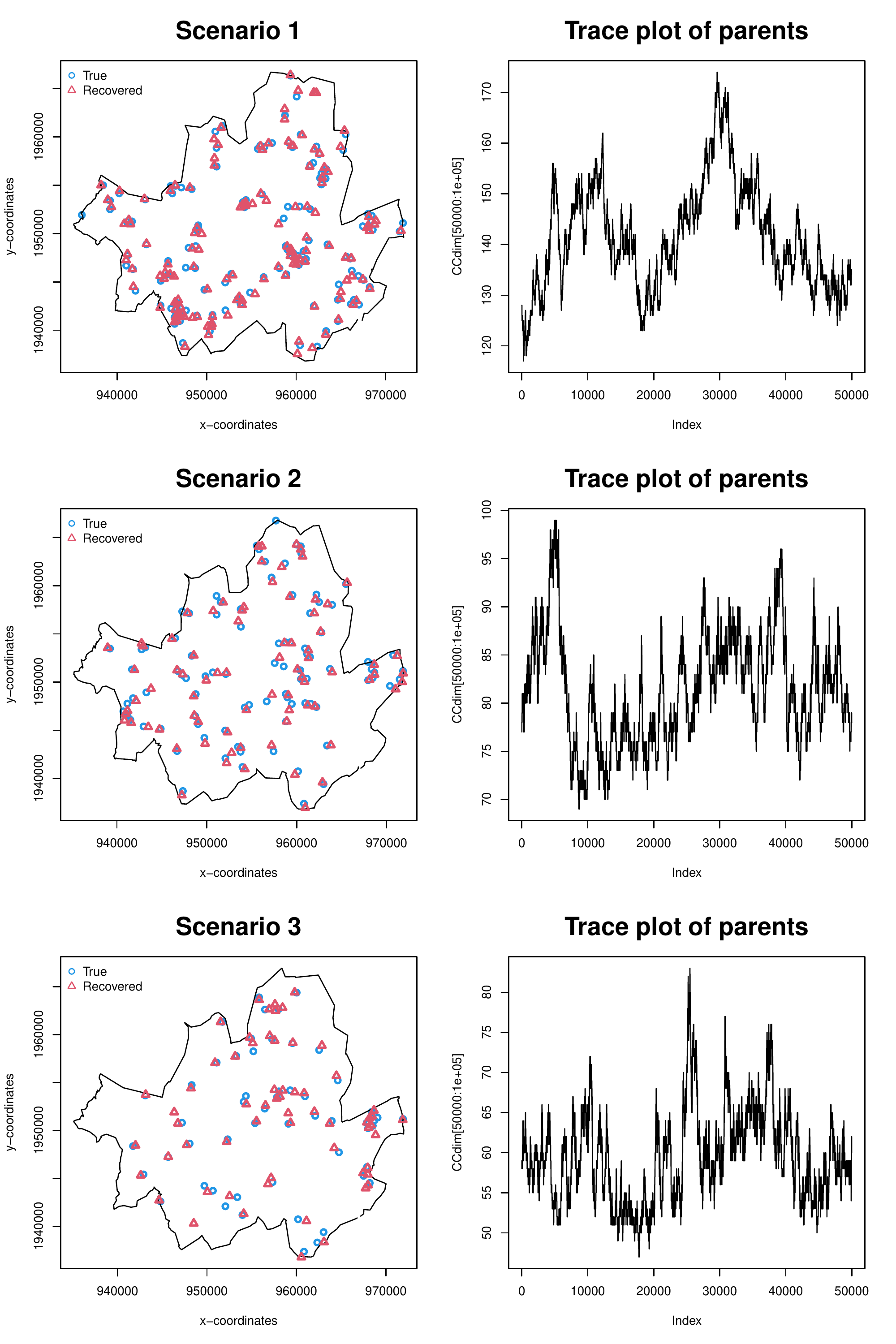}
\end{center}
\caption[]{The left panel compares the true (circle) and recovered (triangle) parent points for different scenarios. The right panel illustrates the trace plots of parents.}
\label{simulpoint}
\end{figure}

Table~\ref{simulest} indicates that the true parameter values used in all scenarios are estimated with reasonable accuracy. We also compare the true and recovered parent points in Figure~\ref{simulpoint}. We obtain the recovered parent points from the last iteration of the MCMC algorithm (i.e., $\mathbf{C}^{(t)}$ in Algorithm~\ref{DMHalg}). We observe that our method can detect spreading events well as there are good agreements between true and fitted points in all scenarios. Trace plots for number of parents also indicate that the MCMC chain from Algorithm~\ref{DMHalg} converges.

\subsection{COVID-19 Contact Tracing Data Analysis}
\label{sec:RealData}


\begin{table}[tt]
\centering
\begin{tabular}{cccccc}
\hline
March 19th  & $\alpha$ & $\omega$ & $\kappa \times 10^7$ & $\theta_{1}$ & $\theta_{2}$ \\
  \hline
Estimates & 5.42 & 357.40 & 1.33 & 1.72 & 547.43\\
95\%HPD & (4.69, 6.11) & (356.26, 359.63) & (0.57, 2.06) & (1.31, 2.21) & (500.00, 638.49) \\
\hline
April 2nd  & $\alpha$ & $\omega$ & $\kappa \times 10^7$ & $\theta_{1}$ & $\theta_{2}$ \\
  \hline
Estimates & 4.61 & 357.44 & 1.00 & 1.92 & 553.22  \\
95\%HPD & (3.88, 5.25) & (356.26, 359.74) & (0.43, 1.71) & (1.37, 2.58) & (500.02, 646.51)\\
\hline
April 15th & $\alpha$ & $\omega$ & $\kappa \times 10^7$ & $\theta_{1}$ & $\theta_{2}$ \\
  \hline
Estimates & 4.08 & 358.55 & 0.56 & 2.08 & 568.40 \\
95\%HPD & (3.24, 4.93) & (356.26, 363.21) & (0.25, 0.88) & (1.34, 2.99) & (500.01, 699.50) \\
\hline
\end{tabular}
\caption{Inference results for real data. 100,000 MCMC samples are generated, discarding the first 50,000 as burn-in.}
\label{realest} 
\end{table}

\begin{figure}[hh]
\begin{center}
\includegraphics[scale = 0.6]{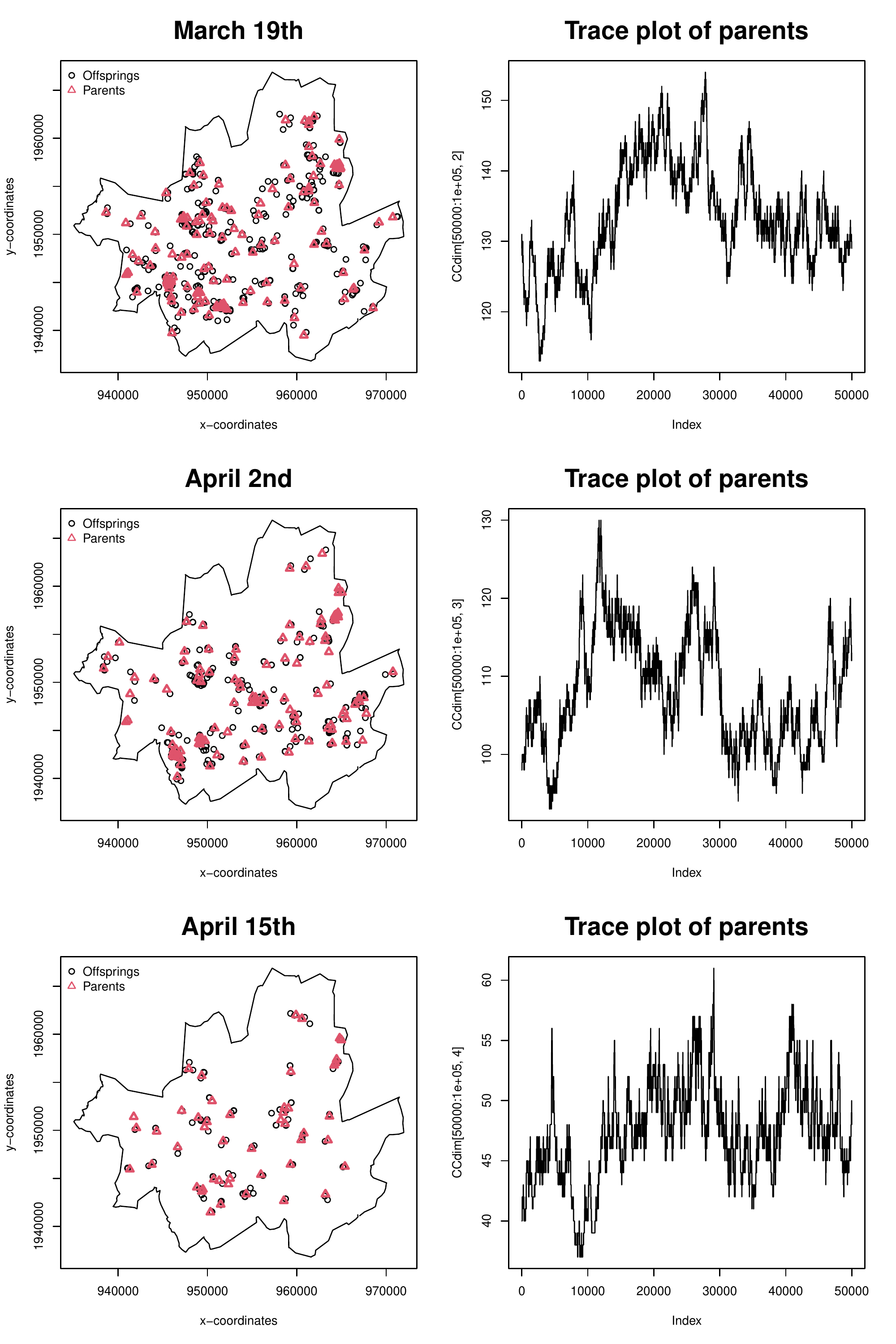}
\end{center}
\caption[]{The left panel compares the true (circle) and recovered (triangle) parent points for real data examples. The right panel illustrates the trace plots of parents.}
\label{realpoint}
\end{figure}

\paragraph{Parameter Interpretation} We now apply our method to the real observational data described in Section \ref{sec:data_description}. The parameter estimation results for time periods that end at March 19th, April 2nd, and April 15th are summarized in Table \ref{realest}. These three consecutive periods can be regarded as severe, moderate, and mild outbreaks, respectively, in terms of the number of patient visits. Parameter estimation results for other time periods are provided in the supplementary material. The average number of offspring points per each parent ($\alpha$) and the intensity for parent ($\kappa$) change as the overall number of visits changes; that is, more overall events lead to higher $\alpha$ and $\kappa$ values. The width of the Gaussian kernel controlled by $\omega$, on the other hand, stays similar. For April 15th, when the total number of visits is relatively small ($n=190$), $\alpha$ takes a lower value as well.

The estimated values for the remaining two parameters $\theta_1$ and $\theta_2$ describe the interaction between parent points and show how clusters of the events are distributed in Seoul. The distance for maximum interaction, determined by $\theta_2$, increases as the number of parents reduces, reflecting a sparser distribution of the parent events. Nevertheless, from the $\theta_1$ estimates (which are significantly greater than 1), parent points in all periods show clear attraction at the location given by $\theta_2$ (around 550m). This implies that different event clusters tend to appear near each other rather than independently.

\begin{figure}[hh]
\begin{center}
\includegraphics[scale = 0.6]{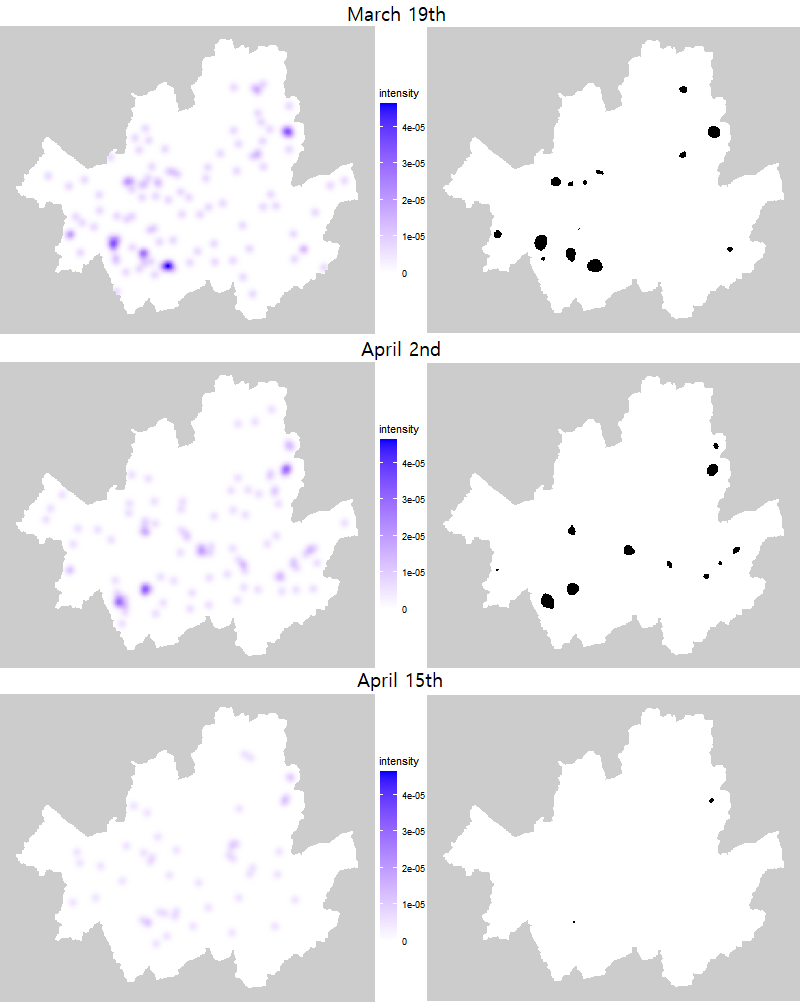}
\end{center}
\caption[]{(Left Column) Intensity map for patient visit events computed by the offspring intensity function in \eqref{eqn:intensity_map}. (Right Column) High risk area with the intensity greater than 1 patient / 1 day 1.427 km$^2$, roughly meaning one patient per day within a "Dong".}
\label{fig:intensity}
\end{figure}

\paragraph{Visualization for COVID-19 Risk} From the estimated values of $\alpha$, $\kappa$, and $\omega$, we can create an intensity map for patient visit events. Since local spreading events cause rapid community transmission, it is essential to avoid the higher risk locations in daily life. Understanding the probabilistic mechanism occurring patient visits and detecting local spreading events (hotspots) are key to manage disease outbreaks. Let $\hat{\alpha}$ and $\hat{\omega}$ be the posterior mean of $\alpha$ and $\omega$ respectively. Given the parent points from the last iteration of the MCMC, denoted as $\hat{c}_1,\cdots,\hat{c}_m$, the spatial intensity function for offspring points is given by 
\begin{equation} \label{eqn:intensity_map}
g(x)=\sum_{i=1}^{m} \alpha k(x-\hat{c}_i,\hat{\omega})
\end{equation}
for any $x \in \mathcal{S}$. This map can be used to identify high-risk areas where the intensity exceeds a certain threshold set by administrative decision. Figure \ref{fig:intensity} illustrates the intensity map $g(x)$ and high-risk areas defined as $g(x)/14 > 1.427$ km$^2$, which corresponds to roughly one patient per $1.427$ km$^2$ each day. Here we divide $g(x)$ by 14 to convert the intensity computed for two weeks into the intensity for one day. The area $1.427$ km$^2$ is the average size of each Dong in Seoul, the smallest administrative district. 

The hotspots shown in Figure~\ref{fig:intensity} present different types of spreading events. For example, the hotspots in Guro-gu (bottom left corner of the map) are caused by a group infection from a call center building around March. The event led to multiple detected confirmed cases. On the other hand, the hotspots in Jungrang-gu (top right corner in the map) are mostly caused by a single patient. This patient moved around multiple places in the local area, which resulted in many close contacts. The former gained considerable media attention, and people became aware of the potential risk in the community. The latter, on the other hand, did not receive media coverage considering privacy protection, and the local community was largely unaware of the risk. Our approach may provide a way to warn local people without revealing information about the specific contagious individual. From Figure~\ref{fig:intensity}, we observe that the overall intensity decreases as time goes, indicating the outbreaks had been controlled around mid-April. In summary, these visualizations allow authorities to identify contagion risk hotspots and help them to set up public health interventions while maintaining privacy protection.

\begin{figure}[hh]
\begin{center}
\includegraphics[scale = 0.6]{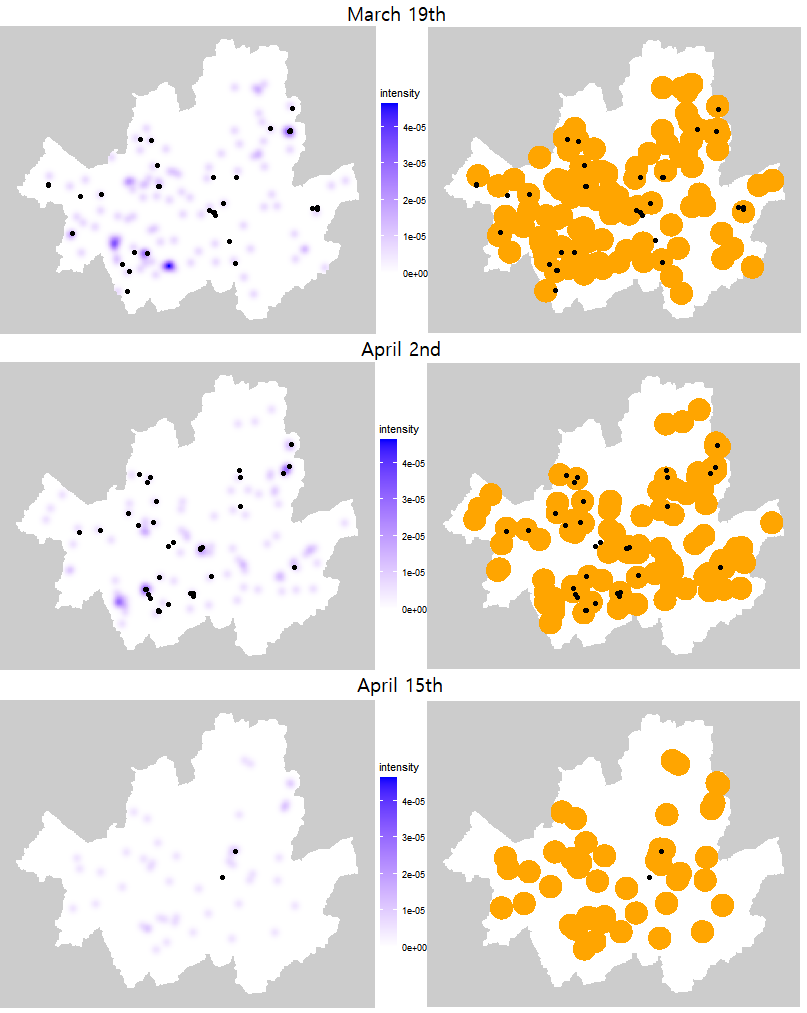}
\end{center}
\caption[]{ The same intensity map as the left column in Figure \ref{fig:intensity} (left column) and the risk boundaries defined as circles, whose centers are the existing sampled parent points $\hat{c}_1,\dots,\hat{c}_m$ and the radii are defined as  $\theta_2+1.96\omega$ (right Column). The black dots in both columns show events observed on the day following the last date of each time interval.}
\label{fig:projection}
\end{figure}

\paragraph{Visualization for Risk Boundaries} Our approach can also provide a way to generate "risk boundaries" within a short time period (such as one day), assuming that the event distribution within that period can be approximated by a model based on the previous two weeks. Such boundaries are potentially useful for issuing public health advisories. The left column of Figure \ref{fig:projection} shows that the patient visits occurred on April 3rd mostly concentrated within the areas with high intensity $g(x)$, given by the existing parent points ($\hat{c}_1,...,\hat{c}_m$), but there are also points located slightly away from high intensity areas. However, there are no points that are clearly far away from the existing high intensity areas either. Therefore,  these points can be viewed as offspring events from new parents that are created near the existing parents.

The right column of Figure \ref{fig:projection} shows the risk boundaries that consider both the occurrence of new parent points and the range of offspring densities created by them. Specifically, risk boundaries are represented as the orange circles, whose centers are the existing parent points $\hat{c}_1,\dots,\hat{c}_m$. The radii are given as $\theta_2+1.96\omega$, the sum of the distance that maximizes the interaction between the parent points ($\theta_2$) and one side of the 95\% interval for the offspring density ($1.96\omega$). As Figure \ref{fig:projection} shows, these risk boundaries indicate how far new events can reach given the existing parent points. The other periods also show qualitatively similar results (Figures 4, 5 in the supplementary material).

\section{Discussion}

In this manuscript, we proposed a novel interaction Neyman-Scott point process for modeling COVID-19 data. Our model can be used to study the spatial patterns of patient visits and detect spreading events in local communities by considering location-based interactions. To address the intractable normalizing functions involved in the posterior, we embed DMH \citep{liang2010double} into our MCMC algorithm. We apply our approach to COVID-19 data sets pertaining to Seoul, and draw meaningful epidemiological conclusions based on parameter estimates. Furthermore, we provide easy-to-interpret disease risk maps from the fitted results. Through simulation studies, we show that our method can provide reasonably accurate parameter estimates and detect true spreading events. 

Here, we focus on developing a new cluster point process model by regarding each patient visit as a realization from the process. Note that there are other spatial models that can be applied to contact tracing data. Examples include dynamic models for movement tracks of animals \citep{russell2016dynamic} and spatial gradient methods for modeling the spread of invasive species \citep{goldstein2019quantifying}. Adopting such methods for studying COVID-19 contact tracing data could provide another interesting epidemiological insight. 

Our method can be useful for planning public health interventions such as deciding the social distancing levels. Social distancing is necessary to prevent the spread of infectious diseases, especially for people in risk groups for severe coronavirus disease. However, raising the social distancing level could adversely affect the local economy. Therefore, it is essential to measure the degree of the COVID-19 risk: {\it{how}} dangerous is the current coronavirus outbreak? Based on our approach, we can provide some guidelines for deciding the appropriate distancing levels to minimize economic damage. We can quantify such risks based on model parameters. For instance, $\alpha$ and $\omega$ estimates provide the expected number of patient visits per spreading event and the levels of spreading events activity (range of spread). From the $\theta_2$ estimate, we can expect the probable location of a new spreading event. Using these parameter estimates, disease control authorities can decide the threshold for social distancing levels. Note that such interpretation is not available from existing cluster point process approaches.

Furthermore, our proposed approach has some advantages considering privacy protection. The disease risk maps allow authorities to alert local communities about disease hotspots without revealing privacy-sensitive information. This is opposite to the current practice in South Korea, where local governments or health authorities release each individual's travel history; although the names of individuals are not revealed, the released information can be used to infer the identity of each patient. The method and ideas developed in this manuscript are generally applicable to other infectious diseases as well.

\section*{Supplementary Material}
The supplementary material available online provides details for birth-death MCMC algorithms and fitted results for other time periods.

\section*{Acknowledgement}
Jaewoo Park was supported by the Yonsei University Research Fund of 2020-22-0501 and the National Research Foundation of Korea (NRF-2020R1C1C1A0100386811). Boseung Choi was supported by the National Research Foundation of Korea (2020R1F1A1A01066082). The Ohio Super Computer Center (OSC) provided part of computational resources for this study. The authors are grateful to Tom{\'a}{\v{s}} Mrkvi{\v{c}}ka for providing useful sample codes and the anonymous reviewers for their careful reading and valuable comments. The authors are also grateful to Korea Spatial Information \& Community Co. Ltd. for their support with data and Sungjae Kim and Eunjin Eom for data collection and organization.

\clearpage
\appendix
\begin{center}
\title{\LARGE\bf Supplementary Material for An Interaction Neyman-Scott Point Process Model for Coronavirus Disease-19}\\~\\
\author{\Large{Jaewoo Park, Won Chang, and Boseung Choi}}
\end{center}

\section{Birth-Death MCMC}

Birth-death samplers \citep{moller2003statistical} are often used to generate point processes given the model parameters. We propose adding a new point (birth), removing an existing point (death), or moving a point with equal probability. To generate an auxiliary variable in DMH, we use a birth-death sampler whose stationary distribution is $f(\mathbf{C}|\kappa,\theta_1,\theta_2)$. Algorithm~\ref{BDMCMCaux} summarizes this step. In our full MCMC algorithm (Algorithm~1 in the manuscript), we update $\mathbf{C}$, given all the model parameters. Here, we use a birth-death sampler whose stationary distribution is $f(\mathbf{X}|\mathbf{C},\alpha,\omega)f(\mathbf{C}|\kappa,\theta_1,\theta_2)$. Algorithm~\ref{BDMCMCC} summarizes this step. 

\begin{algorithm}[htbp]
\caption{Birth-death MCMC for generating the auxiliary variable in DMH}\label{BDMCMCaux}
\begin{algorithmic}
\normalsize
\State \textbf{Given parent parameters $\kappa,\theta_1,\theta_2$ and the point pattern $\mathbf{C}=\lbrace c_1,\cdots,c_m \rbrace$. } \\ 

\State \textbf{Birth step: add a point with probability $1/3$}\\

\State Propose a new point $\xi$ uniformly over the spatial domain $\mathcal{S}$: $\mathbf{C}^{+}=\mathbf{C} \cup \lbrace \xi \rbrace$\\

\State Accept it with probability
$$\alpha=\min\left\lbrace \frac{   f(\mathbf{C}^{+}|\kappa,\theta_{1},\theta_{2})|\mathcal{S}|  }{ f(\mathbf{C}|\kappa,\theta_{1},\theta_{2})(m+1)}, 1 \right\rbrace$$\\

\State \textbf{Death step: remove a point with probability $1/3$}\\

\State Remove an existing point $\eta \in \lbrace c_1,\cdots,c_m \rbrace$: $\mathbf{C}^{-}=\mathbf{C} \setminus \lbrace \eta \rbrace$\\

\State Accept it with probability
$$\alpha=\min\left\lbrace \frac{   f(\mathbf{C}^{-}|\kappa,\theta_{1},\theta_{2})(m-1)  }{ f(\mathbf{C}|\kappa,\theta_{1},\theta_{2})|\mathcal{S}| }, 1 \right\rbrace$$\\

\State \textbf{Move step: move a point with probability $1/3$}\\

\State Propose a new point $\xi$ and remove an existing point $\eta$: $\mathbf{C}^{'}=\mathbf{C} \cup \lbrace \xi \rbrace \setminus \lbrace \eta \rbrace$\\
 
\State Accept it with probability
$$\alpha=\min\left\lbrace \frac{   f(\mathbf{C}^{'}|\kappa,\theta_{1},\theta_{2})  }{ f(\mathbf{C}|\kappa,\theta_{1},\theta_{2}) }, 1 \right\rbrace$$\\

\end{algorithmic}
\end{algorithm}

\begin{algorithm}[htbp]
\caption{Birth-death MCMC for updating $\mathbf{C}$ in the full model}\label{BDMCMCC}
\begin{algorithmic}
\normalsize
\State \textbf{Given parent parameters $\alpha,\omega,\kappa,\theta_1,\theta_2$ and the point pattern $\mathbf{C}=\lbrace c_1,\cdots,c_m \rbrace$. } \\ 

\State \textbf{Birth step: add a point with probability $1/3$}\\

\State Propose a new point $\xi$ uniformly over the spatial domain $\mathcal{S}$: $\mathbf{C}^{+}=\mathbf{C} \cup \lbrace \xi \rbrace$\\

\State Accept it with probability
$$\alpha=\min\left\lbrace \frac{   f(\mathbf{X}|\mathbf{C}^{+},\alpha,\omega)f(\mathbf{C}^{+}|\kappa,\theta_{1},\theta_{2})|\mathcal{S}|  }{ f(\mathbf{X}|\mathbf{C},\alpha,\omega)f(\mathbf{C}|\kappa,\theta_{1},\theta_{2})(m+1) }, 1 \right\rbrace$$\\

\State \textbf{Death step: remove a point with probability $1/3$}\\

\State Remove an existing point $\eta \in \lbrace c_1,\cdots,c_m \rbrace$: $\mathbf{C}^{-}=\mathbf{C} \setminus \lbrace \eta \rbrace$\\

\State Accept it with probability
$$\alpha=\min\left\lbrace \frac{   f(\mathbf{X}|\mathbf{C}^{-},\alpha,\omega)f(\mathbf{C}^{-}|\kappa,\theta_{1},\theta_{2})(m-1)  }{ f(\mathbf{X}|\mathbf{C},\alpha,\omega)f(\mathbf{C}|\kappa,\theta_{1},\theta_{2})|\mathcal{S}| }, 1 \right\rbrace$$\\

\State \textbf{Move step: move a point with probability $1/3$}\\

\State Propose a new point $\xi$ and remove an existing point $\eta$: $\mathbf{C}^{'}=\mathbf{C} \cup \lbrace \xi \rbrace \setminus \lbrace \eta \rbrace$\\
 
\State Accept it with probability
$$\alpha=\min\left\lbrace \frac{   f(\mathbf{X}|\mathbf{C}^{'},\alpha,\omega)f(\mathbf{C}^{'}|\kappa,\theta_{1},\theta_{2})  }{ f(\mathbf{X}|\mathbf{C},\alpha,\omega)f(\mathbf{C}|\kappa,\theta_{1},\theta_{2}) }, 1 \right\rbrace$$\\

\end{algorithmic}
\end{algorithm}

\clearpage
\section{Analysis Results for Different Timelines}

\begin{table}[hh]
\centering
\begin{tabular}{cccccc}
  \hline
March 5th  & $\alpha$ & $\omega$ & $\kappa \times 10^7$ & $\theta_{1}$ & $\theta_{2}$ \\
  \hline
Estimates & 5.02 & 357.98 & 0.89 & 1.75 & 611.67 \\
95\%HPD & (4.40, 5.70) & (356.26, 361.38) & (0.41, 1.43) & (1.25, 2.41) & (500.02, 802.75) \\
\hline
April 30th  & $\alpha$ & $\omega$ & $\kappa \times 10^7$ & $\theta_{1}$ & $\theta_{2}$ \\
  \hline
Estimates & 3.19 & 563.40 & 0.17 &  1.66 & 724.74\\
95\%HPD & (3.00, 3.58) & (402.32, 744.51) & (0.07, 0.29) & (1.00, 2.65) & (500.14, 966.40) \\
\hline
May 14th  & $\alpha$ & $\omega$ & $\kappa \times 10^7$ & $\theta_{1}$ & $\theta_{2}$ \\
  \hline
Estimates & 3.21 & 359.62 & 0.63 & 2.11 & 585.78  \\
95\%HPD & (3.00, 3.56) & (356.26, 366.27) & (0.27, 1.03) & (1.43, 2.97) & (500.00, 745.21) \\
\hline
May 28th  & $\alpha$ & $\omega$ & $\kappa \times 10^7$ & $\theta_{1}$ & $\theta_{2}$ \\
  \hline
Estimates & 5.13 & 357.17 & 1.34 & 1.66 & 538.49\\
95\%HPD & (4.53, 5.79) & (356.26, 358.93) & (0.66, 2.25) & (1.28, 2.13) & (500.01, 607.51) \\
\hline
June 11th & $\alpha$ & $\omega$ & $\kappa \times 10^7$ & $\theta_{1}$ & $\theta_{2}$ \\
  \hline
Estimates & 5.25 & 357.39 & 1.07 &  1.98 & 548.45 \\
95\%HPD & (4.50, 6.07) & (356.27, 359.51) & (0.37, 1.93) & (1.41, 2.66) & (500.02, 644.46) \\
\hline
\end{tabular}
\caption{Inference results for real data. 100,000 MCMC samples are generated, discarding the first 50,000 as burn-in.}
\label{realest2} 
\end{table}

\begin{figure}[tt]
\begin{center}
\includegraphics[scale = 0.6]{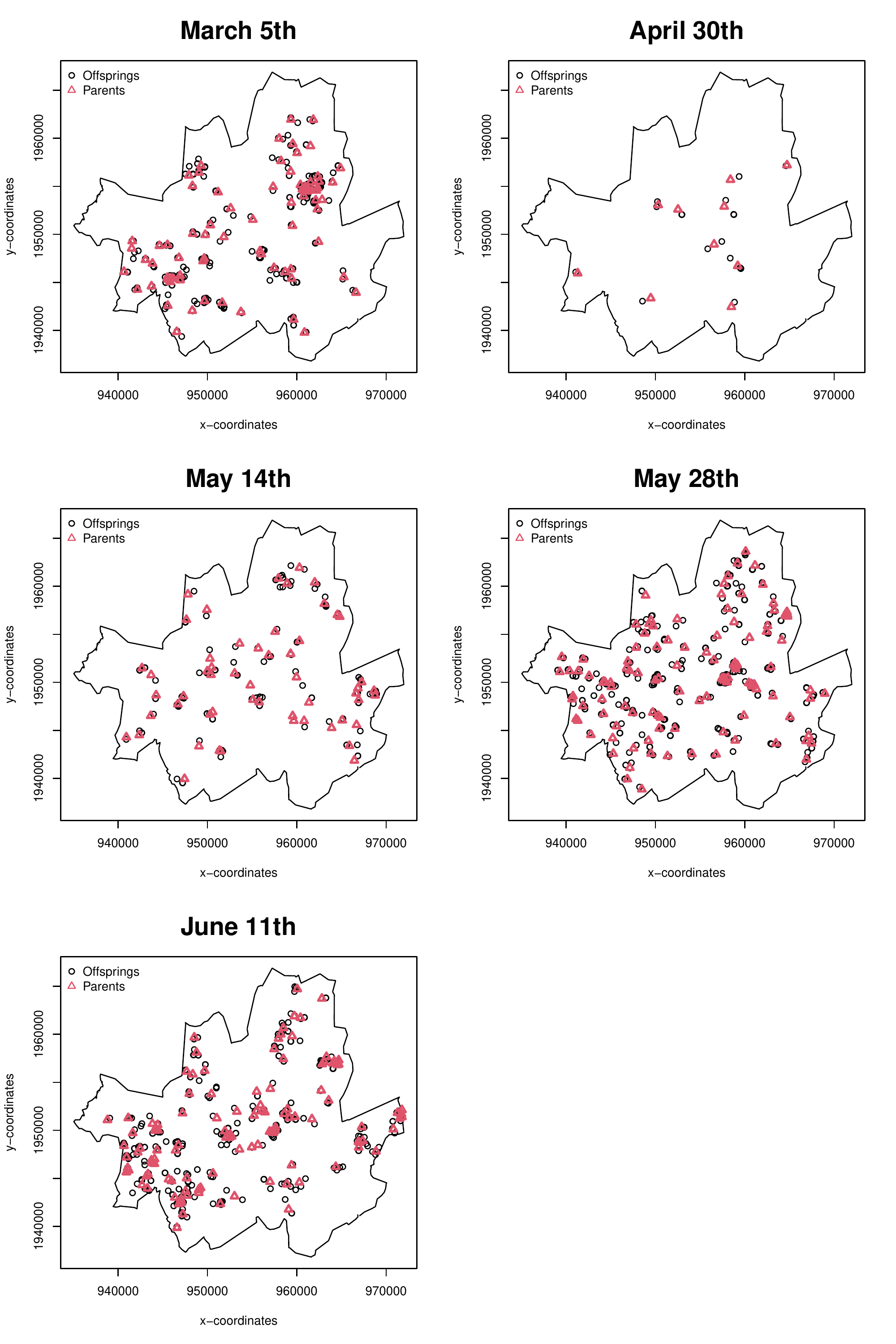}
\end{center}
\caption[]{Recovered parent points for the real data example. Circles indicate offsprings and triangles indicate recovered parent points.}
\label{realpoint2}
\end{figure}

\begin{figure}[tt]
\begin{center}
\includegraphics[scale = 0.65]{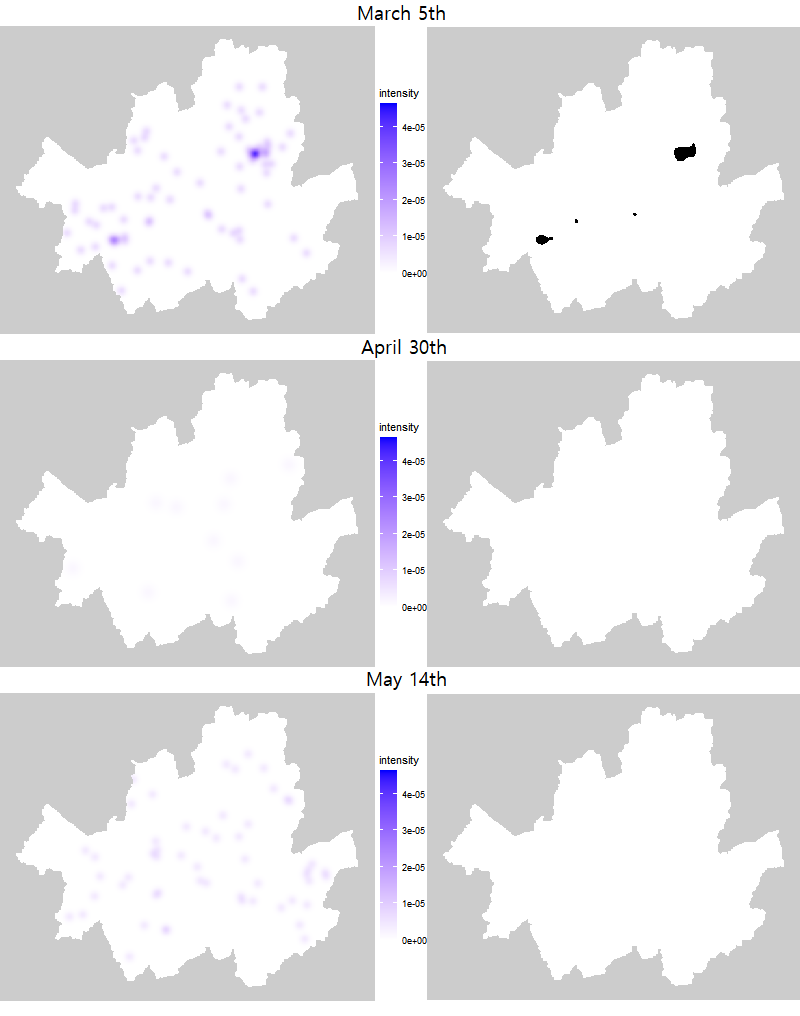}
\end{center}
\caption[]{(Left column) Intensity map of the patient visit events. (Right column) High-risk area with intensity greater than 1 patient/1 day 1.427 km$^2$, which roughly translates to one patient per day within a "Dong", the smallest administrative unit.}
\label{realinten1}
\end{figure}

\begin{figure}[tt]
\begin{center}
\includegraphics[scale = 0.65]{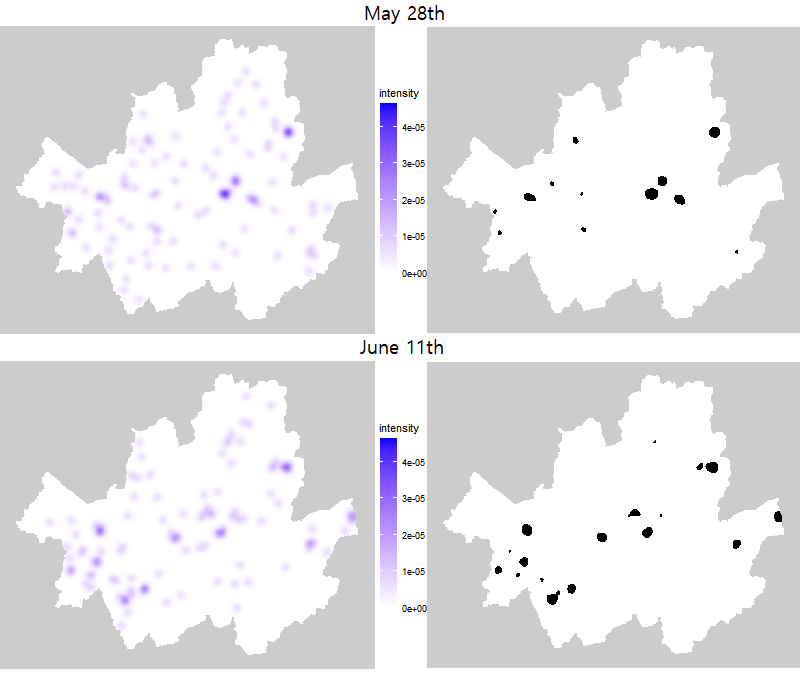}
\end{center}
\caption[]{(Left column) Intensity map of the patient visit events. (Right column) High-risk area with intensity greater than 1 patient/1 day 1.427 km$^2$, which roughly translates to one patient per day within a "Dong", the smallest administrative unit.}
\label{realinten2}
\end{figure}

\begin{figure}[tt]
\begin{center}
\includegraphics[scale = 0.65]{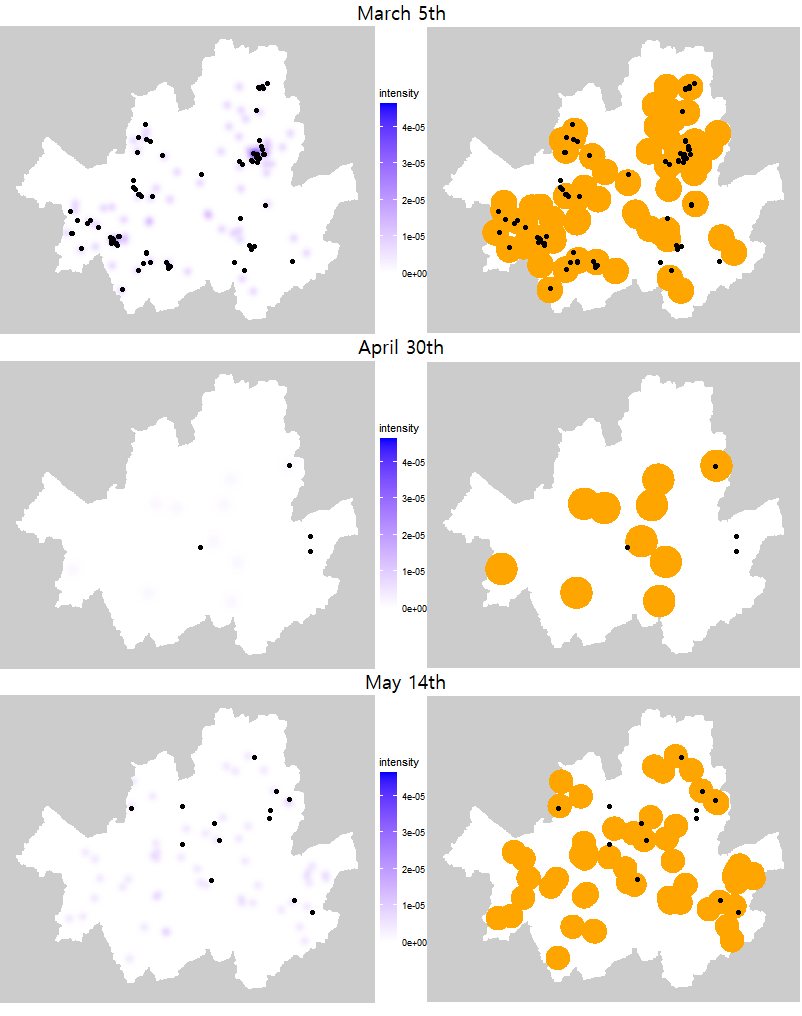}
\end{center}
\caption[]{The intensity map (left column) and risk boundaries defined as circles, whose centers are the existing sampled parent points $\hat{c}_1,\dots,\hat{c}_m$ and the radii are defined as  $\theta_2+1.96\omega$ (right column). The black dots in both columns show the observed events during the day following the last date of each time interval.}
\label{realinten3}
\end{figure}

\begin{figure}[tt]
\begin{center}
\includegraphics[scale = 0.65]{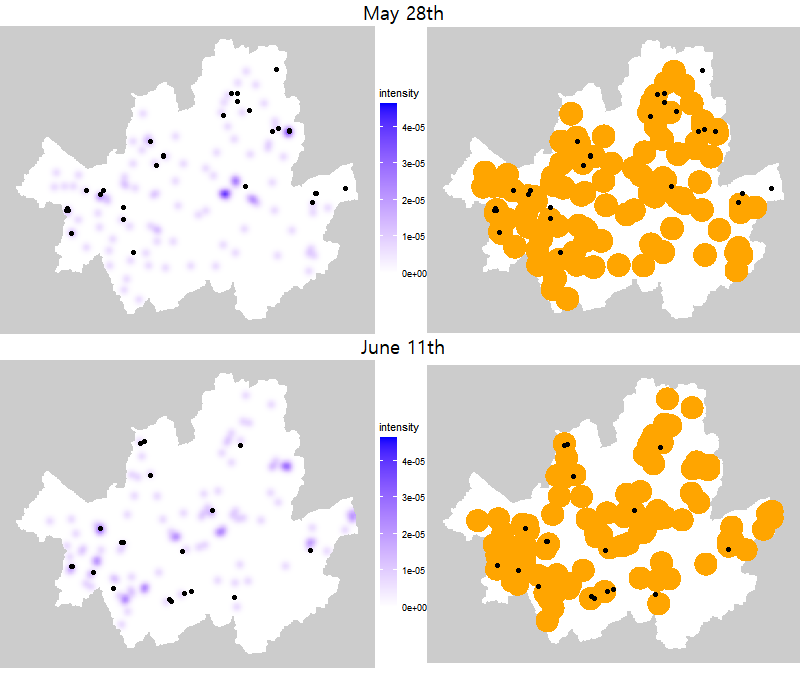}
\end{center}
\caption[]{The intensity map (left column) and risk boundaries defined as circles, whose centers are the existing sampled parent points $\hat{c}_1,\dots,\hat{c}_m$, and the radii are defined as  $\theta_2+1.96\omega$ (right column). The black dots in both columns show the observed events during the day following the last date of each time interval.}
\label{realinten4}
\end{figure}


\clearpage
\bibliography{Reference}
\end{document}